\documentclass[letter]{aa}  
\usepackage{graphicx}
\usepackage{txfonts}
\usepackage{natbib}
\usepackage{xcolor}
\usepackage[normalem]{ulem}

\begin{document} 

   \title{Radio interferometric imaging of RS Oph bipolar ejecta\\
   for the 2021 nova outburst} 
   \titlerunning{Radio imaging of RS Oph in 2021}
   \authorrunning{U. Munari et al.}
   \author{U. Munari
          \inst{1},
          M. Giroletti
          \inst{2},
          B. Marcote
          \inst{3},
          T.~J. O'Brien
          \inst{4},
          P. Veres
          \inst{5},
          J. Yang
          \inst{6},
          D.~R.~A. Williams,
          \inst{4}
          \and
          P. Woudt
          \inst{7}
          }
   \institute{  INAF Osservatorio Astronomico di Padova, 36012 Asiago (VI), Italy,\\
              \email{ulisse.munari@inaf.it}
         \and
                INAF Istituto di Radioastronomia, via Gobetti 101, 40129, Bologna, Italy  
         \and   
                Joint Institute for VLBI ERIC, Oude Hoogeveensedijk 4, 7991~PD Dwingeloo, The Netherlands
         \and
                Jodrell Bank Centre for Astrophysics, School of Physics and Astronomy, University of Manchester, Manchester, M13 9PL, UK
         \and
                Center for Space Plasma and Aeronomic Research (CSPAR), University of Alabama in Huntsville, Huntsville, AL 35899, USA
         \and
                Dept. of Space, Earth and Environment, Chalmers University of Technology, Onsala Space Observatory, SE-43992 Onsala, Sweden
         \and
                Department of Astronomy University of Cape Town, Private Bag X3, Rondebosch 7701, South Africa
             }

   \titlerunning{Radio imaging of RS Oph 2021}   

   \date{Received September 15, 1996; accepted March 16, 1997}

  \abstract{
  The recurrent nova and symbiotic binary RS Oph erupted again in August
  2021 for its eighth known outburst.  We observed RS~Oph 34 days after the
  outburst at 5~GHz with the European VLBI Network (EVN).  The radio image
  is elongated over the east--west direction for a total extension of about
  90 mas (or about 240 AU at the {\it Gaia} DR3 distance
  $d=2.68_{-0.15}^{+0.17}$ kpc), and shows a bright and compact central
  component coincident with the {\it Gaia} astrometric position, and two
  lobes east and west of it, expanding perpendicular to the orbital plane. 
  By comparing with the evolution of emission-line profiles on optical
  spectra, we found the leading edge of the lobes to be expanding at
  $\sim$7550~km\,s$^{-1}$, and $i$=54$^\circ$ as the orbital inclination of
  the binary.  The 2021 radio structure is remarkably similar to that
  observed following the 2006 eruption.  The obscuring role of the density
  enhancement on the orbital plane (DEOP) is discussed in connection to the
  time-dependent visibility of the receding lobe in the background to the
  DEOP, and the origin of the triple-peaked profiles is traced to the ring
  structure formed by the nova ejecta impacting the DEOP.}
   \maketitle

\section{Introduction}

The recurrent nova and symbiotic binary RS Oph underwent a new outburst in
August 2021, following previous outbursts recorded in 1898, 1933, 1958, 1967,
1985, and 2006, with a further probable eruption in 1945 that was
partly obscured by the seasonal gap \citep{2011MNRAS.414.2195A}.  According
to \citet{2021arXiv210901101M}, the eruption started on ($t_{\rm ini}$) 2021 Aug
08.50($\pm$0.01) UT and the passage through maximum optical brightness ($t_{\rm
max}$) occurred on 2021 Aug 09.58 ($\pm$0.05) UT; $t_{\rm ini}$ will constitute
our reference epoch throughout this paper.

A nova eruption is rather violent, with 10$^{-6}$ to 10$^{-4}$~M$_\odot$ of
material being expelled at several 10$^3$ km\,s$^{-1}$ as a consequence of
thermonuclear runaway (TNR) igniting in the accreted and electron-degenerate
shell of a white dwarf (WD).  The role played by the powerful shocks that
develop in the resulting (multi-component) outflow is being increasingly
appreciated \citep[e.g.][]{2020ApJ...905...62A, 2021ApJS..257...49C,
2021ARA&A..59..391C}, and even more so for novae that erupt within symbiotic
binaries \citep[see][for a recent review]{2019arXiv190901389M}, in which the
pre-existing wind from the red giant (RG) companion is dense and massive
enough to provide a decelerating medium for the nova ejecta that slam into
it.

The RS Oph outburst of 2021 triggered widespread interest and was followed
across the entire electromagnetic spectrum with meticulous observations; so
far, only a very small proportion of the information stemming from these
observations has been published.  \citet{2022ApJ...935...44C} presented the
outburst evolution in $\gamma$-rays at energies $>$0.1~GeV as recorded by
{\it Fermi}-LAT, showing a broad peak (width $\sim$1 day) centred perfectly
on the optical maximum.  \citet{2022Sci...376...77H} and
\citet{2022NatAs...6..689A} reported the detection of RS Oph at the even
higher energies probed by the Cherenkov telescopes, peaking at day +3.5. 
Dense monitoring of RS Oph in the X-rays was performed by the {\it Swift}
satellite, which was modelled and compared to similarly extensive
observations obtained during the 2006 eruption by
\citet{2022MNRAS.514.1557P}.  The shock-powered, hard X-rays reached maximum
intensity on day +5 and behaved very similarly to those observed during the
2006 event, as did the photometric light curve at optical wavelengths.  The
supersoft X-ray component emerged between days +20 and +25, when the ejected
circumstellar material had become sufficiently diluted to clear the view to
the central binary.  The spectral evolution has also been a close replica of
previous eruptions: at optical wavelengths it was documented at high
resolution and daily cadence by \citet{2021arXiv210901101M,
2022arXiv220301378M}, and revealed the initial flash-ionization by the TNR
of the RG wind, its recombination following an $e$-folding time of $\sim$60
hours, and the progressive deceleration of the nova ejecta plowing through
the pre-existing RG wind.

\begin{table*}
\caption{Results of fitting the delta and ring components as shown in the
right panel of Figure~\ref{fig:fig1}.}
\label{tab:modelfit}
\centering                              
\begin{tabular}{cccccc}    
\hline\hline                
&&\\
Component & Type & RA & Dec & Radius & Flux density \\
&&\\
\hline                              
&&\\
{$\it CC$} & delta & 17h50m13.1608s & $-$06d42\arcmin28.6082\arcsec
 & - - & 0.91 mJy \\
{$\it WL$} & ring & 17h50m13.1596s & $-$06d42\arcmin28.6053\arcsec
 & 22.5 mas & 5.6 mJy \\
&&\\
\hline                  
\end{tabular}
\end{table*}

   \begin{figure*}
   \includegraphics[width=0.58\textwidth]{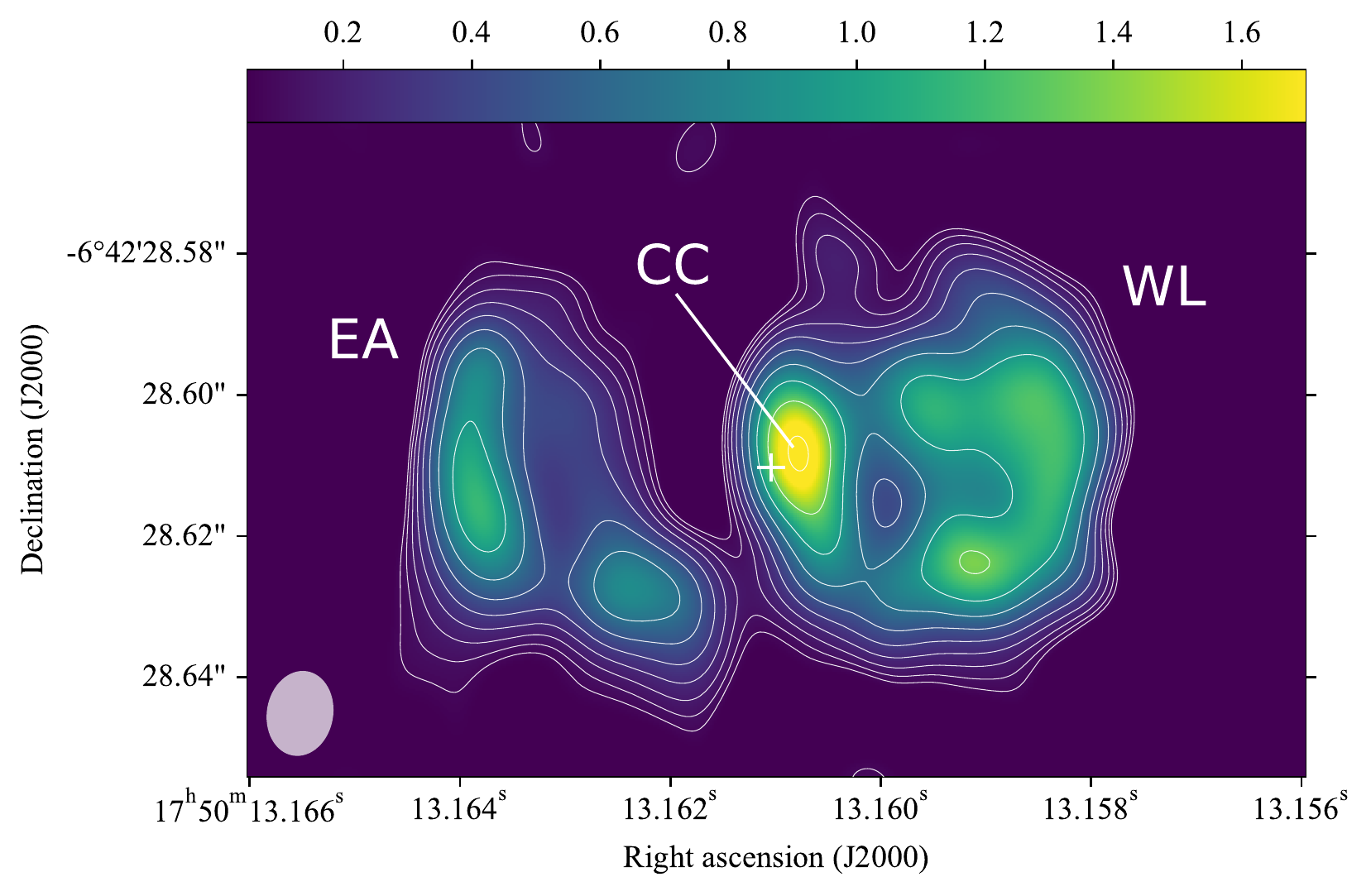}
   \includegraphics[width=0.42\textwidth]{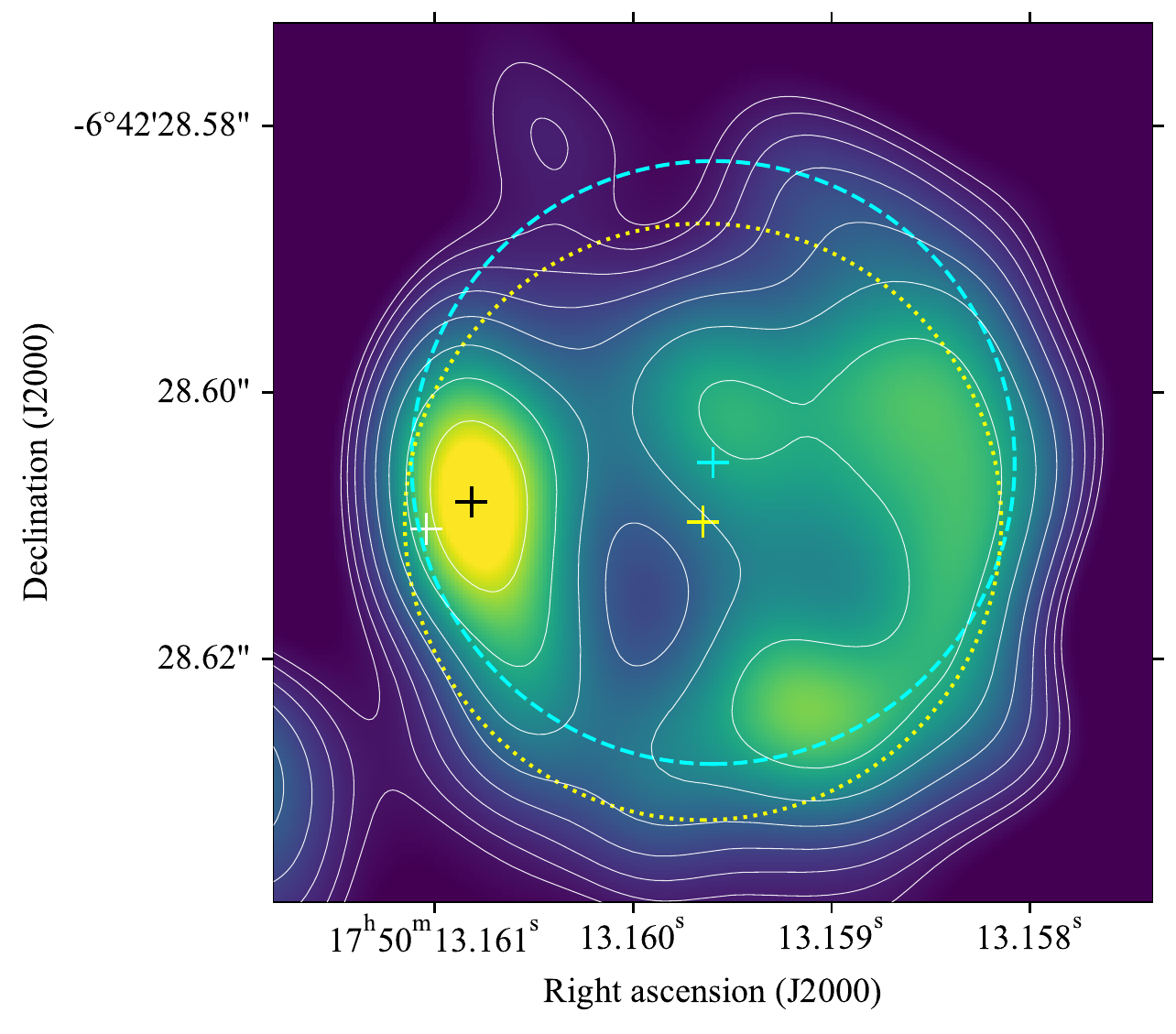}
      \caption{5~GHz EVN+\textit{e}-MERLIN image of RS Oph on 2021 Sept 11,
      34.3 days from $t_{\rm ini}$.  {\it Left}: full scale image.  The beam
      size is shown by the ellipse in the bottom left corner (9.4\,mas
      $\times$ 12.1\,mas in p.a.\ $-8.5^\circ$).  The white cross represents
      the position and uncertainty (magnified by a factor of 50 for
      visibility) of RS Oph in {\it Gaia} DR3, moved to day +0 by
      application of proper motions.  {\it CC} refers to the compact
      component close to the {\it Gaia} position, {\it WL} to the western
      round lobe, and {\it EA} is the arc structure to the east.  The colour
      scale and contours represent total intensity emission of the cleaned
      image, starting at $\pm5\sigma$ ($\pm$0.25 mJy\,beam$^{-1}$) and
      increasing by steps of $\sqrt{2}$.  {\it Right}: zoom onto the {\it
      WL} component, with the best-fit ring and centre from 2021 data
      overlaid as a cyan dashed line, and the best-fit ring and centre for
      2006 from \citet{2008ApJ...688..559R} shown with the yellow dotted
      line, ported to the same 2021 epoch by application of {\it Gaia} DR3
      proper motions.  The black cross marks the delta component accounting
      for most of the emission from {\it CC}.  The colour scale is the same
      as in the left panel, and contours start at $8\sigma$ for the sake of
      clarity.}
         \label{fig:fig1}
   \end{figure*}

Preliminary reports of radio observations made during the 2021 outburst of
RS Oph have been issued by \citet{2021ATel14849....1W},
\citet{2021ATel14886....1S}, \citet{2021ATel14899....1N}, and
\citet{2021ATel14908....1P}.  These observations explored the frequency
range from 0.8 to 35 GHz, started as early as day +2, and recorded an
increase in flux with passing days up to the latest observing epoch reported
(day +28).  They also recorded changes in the spectral index from an initial
$\alpha$$\approx$+0.8 to a later $\alpha$$\approx$$-$0.3.  The initial
inverted spectrum was attributed to external free-free absorption or to
synchrotron self-absorption within the radio-emitting region, while the
flatter slope and deviations from a simple power law at later epochs (when
the absorption was assumed to have mostly cleared) were attributed to
inhomogeneities in the emitting region or to partial obscuration by an
inhomogeneous absorbing screen.  The derived brightness temperatures,
ranging from 10$^6$~K to 10$^8$~K, indicate a non-thermal origin for the
radio emission.

In this Letter we present a European VLBI Network (EVN) radio map of RS Oph
from observations made on day +34 at 5 GHz, which is part of a longer
monitoring campaign, the details and comprehensive analysis of which will be
presented in a dedicated forthcoming publication (M.  Giroletti et al., in
prep.).

\section{Radio observations}

We observed RS\,Oph with the EVN ---including \textit{e}-MERLIN
outstations--- on 2021 September 11, for 8 hours starting at 14.30 UT (day
+34.3 at mid run).  The observing waveband started at 4.81\,GHz, with a
bandwidth of 256\,MHz, divided into 8$\times$32\,MHz-wide sub-bands, each
recorded in dual-polarisation, for a total data rate of 2\,Gbps.  The
participating telescopes were: Jodrell Bank (Mark2), Westerbork, Effelsberg,
Medicina, Shanghai (25\,m), Toru\'n, Hartebeesthoek, Irbene, Yebes, Svetloe,
Zelenchukskaya, Badary, and the \textit{e}-MERLIN outstations in Cambridge,
Darnhall, Defford, Knockin, and Pickmere.  The shortest baseline is
Darnhall-Pickmere (16\,km, corresponding to 650\,mas angular scale), and the
longest is Shanghai-Hartebeesthoek (10161\,km, corresponding to 1.0\,mas
angular scale).  Data were electronically transferred to the Joint Institute
for VLBI~ERIC (JIVE, The Netherlands) and correlated in real time with the
SFXC software correlator \citep{2015ExA....39..259K}.

The observations followed a standard phase-referencing scheme, with four-minute
scans on RS\,Oph bracketed by 40-second calibrator scans on the source
J1745$-$0753 at a separation of 1$^\circ$.67.  The initial data reduction
was carried out using the latest version of the EVN pipeline based on
ParselTongue and AIPS.  The calibrated data were then downloaded and
inspected with Difmap.  Editing, cleaning, and self-calibration were carried
out interactively, switching between uniform and natural weights to optimise
the contributions of both the short and the long baselines.  We also checked
the reliability of the self-calibration solutions with task {\tt CALIB} in
AIPS, which confirmed smooth variations in phase solutions.

The left panel of Figure~\ref{fig:fig1} shows the final EVN image.  It was
obtained in Difmap with natural weights and a Gaussian taper of 0.5 at 10
M$\lambda$, providing a restoring beam of 9.4\,mas $\times$ 12.1\,mas in
p.a.\ $-8.5^\circ$.  The off-source r.m.s.\ noise is
50\,$\mu$Jy\,beam$^{-1}$.  The total cleaned flux density is 20 mJy, which
corresponds to a monochromatic luminosity of 1.7$\times
10^{13}$\,W\,Hz$^{-1}$.

\section{Discussion}

\subsection{Radio components}

The radio emission of RS Oph in Figure~\ref{fig:fig1} is elongated over the
east--west direction for a total extension of about 90 mas (or about 240 AU
at the {\it Gaia} DR3 distance $d=2.68_{-0.15}^{+0.17}$ kpc).  A compact
central component ({\it CC}) lies within a few mas of the {\it Gaia} DR3
position \citep{2022yCat.1355....0G} extrapolated to the observing epoch. 
Emission is present on either side of the {\it CC}.  This emission is brighter
on the western side, forming a circular lobe ({\it WL}), similarly to what
was observed in the 2006 explosion \citep{2006Natur.442..279O}.  Fainter
emission is present to the east of {\it CC}, in the shape of an arc ({\it
EA}).

   \begin{figure}[!t]
   \centering
   \includegraphics[width=9cm]{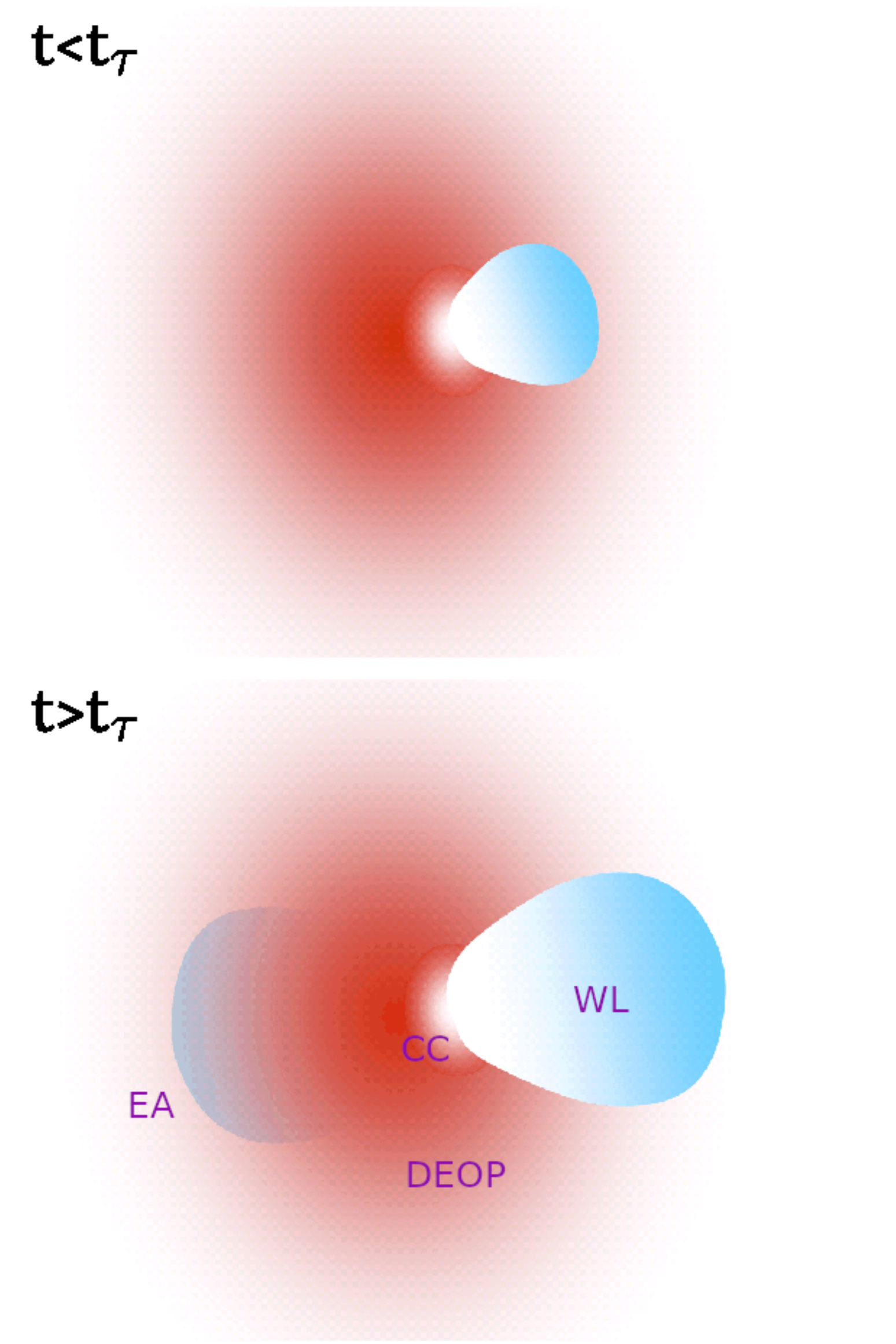}
      \caption{Simple sketch (not to scale) of the expanding and bipolar
               arrangement of RS Oph ejecta.  The orbital plane is vertical
               and the viewing angle matches the
               $i$$\sim$$50^\circ$$\pm$$5^\circ$ inclination inferred from
               the radial velocity orbit and assumed masses for the RG and
               WD.  For the definition of $t_\tau$ (about day +20) and the
               meaning of the other labels, see sect.  3.2.  The lower panel
               represents the view to RS Oph at the time of the day +34.3
               EVN observation, when the leading edge of the receding lobe
               ({\it EA}) has moved past the projected distance from the
               centre where the free-free opacity of the DEOP in the foreground
               drops to $\tau_\nu$=1.}
         \label{fig:fig3}
   \end{figure}

Preliminary inspection of the radio images of RS Oph we collected with EVN
at other epochs shows that the {\it WL} brightens and expands with time.  In
order to provide a simple yet quantitative representation of the {\it CC}
and {\it WL}, we processed the visibilities with a mixed approach. 
From the final dataset, we removed all the clean components from {\it CC}
and those on the outer front of {\it WL}, preserving only those in {\it EA}
and in the inner $\sim20$ mas of {\it WL}.  We then carried out an
interactive model fitting adding a delta component at the central position
and a circular ring on the western residuals.  The results of this fit are
shown in Table~\ref{tab:modelfit}.  The resulting ring and its centre are
shown in cyan overlaid on the clean image in the right panel of
Fig.~\ref{fig:fig1}.

A similar expanding ring was observed by \citet{2006Natur.442..279O,
2008ASPC..401..239O}, \citet{2008ApJ...685L.137S}, and
\citet{2008ApJ...688..559R} in their VLBI, VLBA, EVN, and MERLIN multi-epoch
imaging of RS Oph during the 2006 outburst.  On day +26.8 of the 2006
outburst, \citet{2008ApJ...688..559R} measured the radius of the ring at 5
GHz, finding 17.5($\pm$0.5) mas, and found its centre laying at (J2000)
17$^h$50$^m$13.1583$^s$, $-$6$^\circ$42$^\prime$28.517$^{\prime\prime}$
($\pm$1 mas).  In the right panel of Figure~\ref{fig:fig1}, we overplot the
centre and ring found by these latter authors in yellow (with its radius
linearly increased to match our day +34.3 epoch).  The agreement with the
2021 values is remarkable.

\subsection{A model for the radio emission of RS Oph}

The basic components of our 2021 radio map in Figure~\ref{fig:fig1} were
also present in 2006.  The inaccurate ground-based astrometry in 2006
placed RS Oph $\sim$100~mas away from the radio structure, which did not
help in modelling the observations.  While a bipolar structure was considered
possible by \citet{2006Natur.442..279O}, \citet{2008ApJ...688..559R}
interpreted {\it WL} as the outer front of the shocked ejecta, expanding in
a spherical arrangement away from RS Oph located at the centre, with {\it
EA} being a polar jet leaving the central binary at a much greater  velocity
than the spherical ejecta, while {\it CC} was considered part of an
east--west asymmetry in the brightness distribution of {\it WL}.

Thanks to the superb accuracy of the {\it Gaia} DR3 astrometric position for
RS Oph, we are now able to propose a different model for the radio emission
of RS Oph that abandons spherical symmetry for the ejecta in favour of a
clean bipolar arrangement.  The simple sketch in Figure~\ref{fig:fig3}
outlines the basic spatial arrangement of our model and its temporal
evolution, for a viewing angle matching the
$i$$\sim$$50^\circ$$\pm$$5^\circ$ orbital inclination inferred from orbital
solutions of the radial velocity curve of the RG, and the assumed mass of
the RG and the WD \citep{1996AJ....111.2090D, 2000AJ....119.1375F,
2009A&A...497..815B}.

In addition to a massive accretion disk (AD) around the WD, 3D
hydrodynamical simulations of symbiotic stars show that much of the mass
loss from the RG is gravitationally focused by the companion towards the
orbital plane, creating a strong and disk-like density enhancement with the
binary at its centre \citep[e.g.][]{2008A&A...484L...9W,
2012MNRAS.419.2329O, 2013A&A...551A..37M, 2015ApJ...806...27P,
2016MNRAS.457..822B, 2022ApJ...931..142L}.  In Figure~\ref{fig:fig3}, the
density enhancement on the orbital plane (DEOP) is represented by the
diffuse and reddish component.

During a nova eruption, the combined effect of the AD and the DEOP is to
confine the ejecta primarily within a bipolar structure, which expands
perpendicular to the orbital plane.  The non-thermal radio emission
originates at the shock interface between the fast expanding lobes and the
pre-existing slow wind of the RG.  The {\it WL} component in
Figure~\ref{fig:fig1} is the projection on the plane of the sky of the lobe
moving towards us, in the foreground to the DEOP.  {\it EA} is the visible
outer portion of the lobe expanding opposite the {\it WL}, moving away from
us and located {behind the} DEOP.

The projected surface density of the DEOP, and probably also its ionisation
degree, radially declines moving away from the central binary, which implies
that the free-free opacity exerted by the DEOP also reduces as a function of
the increasing projected distance from its centre.  For its synchrotron
radiation (of frequency $\nu$) to be able to reach us, the {\it EA}
component has to move to a projected distance from the central binary
($\theta_\nu$) that is large enough to cross the DEOP where the optical
depth turns $\tau_\nu$$<$1.

As sketched in Figure~\ref{fig:fig3}, at epochs $t$$<$$t_\tau$ (where
$t_\tau$ designates the time {\it EA} needs to expand to $\theta_\nu$ where
$\tau_\nu$$=$1), the radio emission from {\it EA} is completely absorbed by
the DEOP, and its presence therefore goes unrevealed.  Observations on days
+13.8 by \citet{2006Natur.442..279O} and +20.8 by
\citet{2008ApJ...688..559R} pertain to this phase.  Conversely, at later
$t$$>$$t_\tau$ epochs, the emission from {\it EA} becomes visible, first
from its leading edge and then from the rest, progressively growing in flux
density and angular extension as more and more of {\it EA} moves beyond the
$\tau_\nu$$=$1 position.  Epochs +21.5 and +28.7 of
\citet{2006Natur.442..279O}, +26.8 of \citet{2008ApJ...688..559R}, +49 of
\citet{2008ASPC..401..239O}, and +51 of \citet{2008ApJ...685L.137S} trace
this emersion of {\it EA} into visibility.  Ultimately, when the whole of
{\it EA} has moved past $\theta_\nu$, it will appear similar in shape and
angular extension to {\it WL}, and at equal angular distance from the RS Oph
astrometric position, a scenario that appears to be consistent with the day
+63 radio image of RS Oph presented by \citet{2008ASPC..401..239O}.

In connection to the obscuration exerted by the DEOP, it is useful to note
that \citet{2008ApJ...688..559R} computed the free-free opacity at 5~GHz
predicted for the fully ionised $r^{-2}$ wind from the RG in RS Oph as a
function of the observed projected distance.  Assuming a mass-loss rate
$\dot{M}$=1$\times$10$^{-6}$~M$_\odot$\,yr$^{-1}$ and a
$v_w$=20~km\,s$^{-1}$ wind velocity, these latter authors found the
$\tau_\nu$=1 threshold being crossed at a projected $\sim$24 mas (65~AU)
distance.  A lower (higher) mass-loss rate would shorten (widen) this
separation.  Although the actual $\dot{M}$ of RG or the density profile and
ionisation degree of the DEOP may be different, this value is comparable to
the distance from the {\it Gaia} astrometric position at which the inner
edge of {\it EA} is located in Figure~\ref{fig:fig1}.

We interpret the {\it CC}  as the interface between the DEOP and the
impacting fast ejecta.  Its much higher density turns the DEOP into a
decelerating medium that is far more efficient than the diffuse wind of RG:
a lot of kinetic energy is transformed within a small volume of space,
leading to the bright non-thermal emission spike of Figure~\ref{fig:fig1}.

This proposed spatial arrangement for the radio-emitting regions of RS Oph
is broadly consistent with the [OIII] imaging obtained by {\it HST} at 155
and 449 days after the 2006 outburst, which has been discussed by
\citet{2009ApJ...703.1955R}.  The double-ring structure observed by these
latter authors on day +155 has an east–west orientation with a total
(peak-to-peak) extent of 360($\pm$30)~mas in the plane of the sky,
corresponding to a (constant) expansion rate of
1.16($\pm$0.1)~mas\,day$^{-1}$.  Projecting it back to our day +34.3
observing epoch, the corresponding total extension would be $\sim$80 mas.

\subsection{Comparison with radial velocities}

   \begin{figure}
   \centering
   \includegraphics[width=9cm]{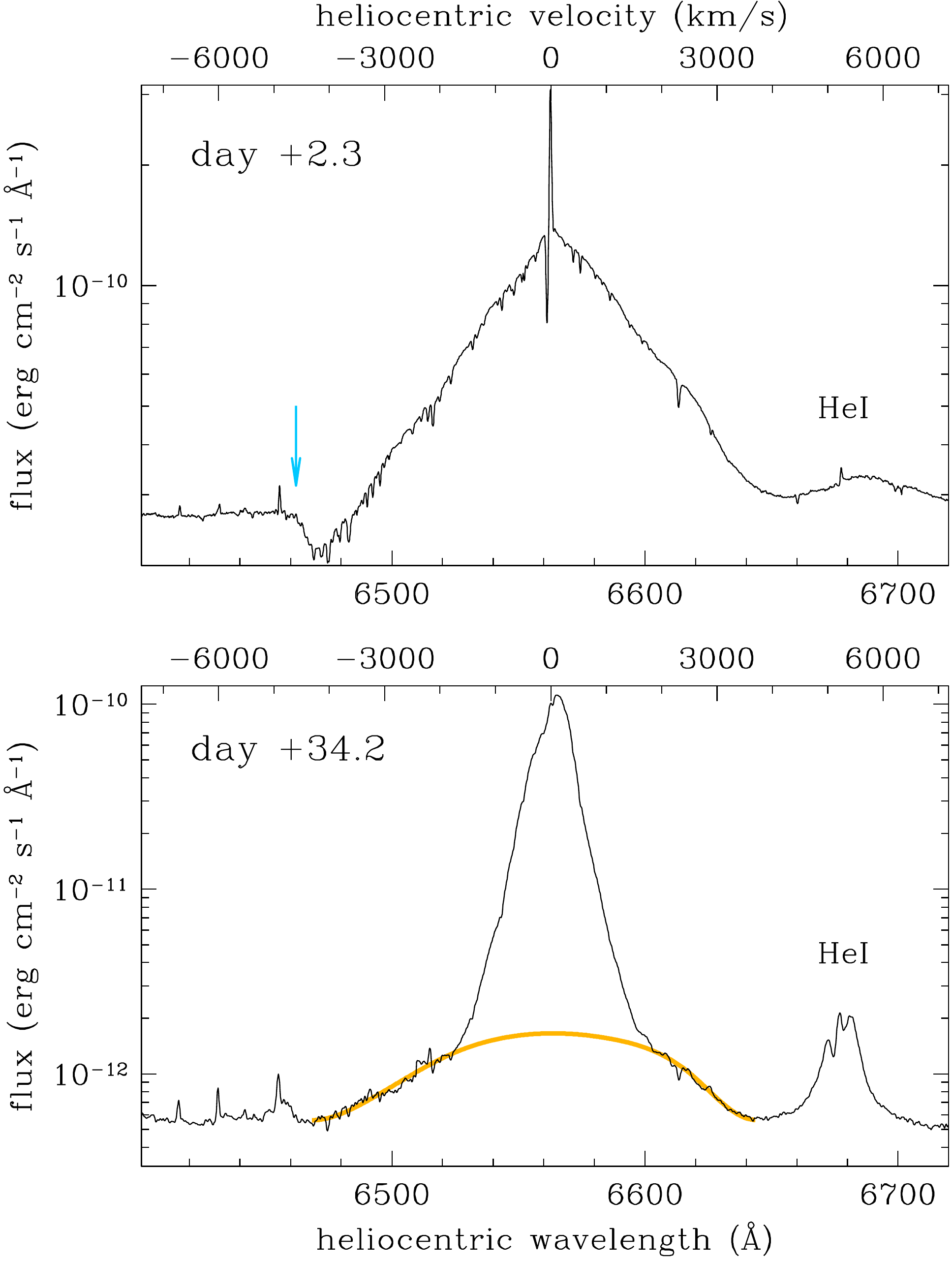}
      \caption{Sample H$\alpha$ profiles of RS Oph \citep[from data
               in][]{2022arXiv220301378M} early in the outburst and at the
               time of our EVN radio imaging (see sect 3.3 for details). 
               Heliocentric radial velocities are given at the top of the
               panels.  The arrow in the top panel marks the
               $-$4700~km\,s$^{-1}$ terminal velocity of the P-Cyg
               absorption, while the orange line in the lower panel is a fit to
               the broad pedestal extending up to $-$4200~km\,s$^{-1}$.}
         \label{fig:fig4}
   \end{figure}

To gain insight into the space velocity of the ejecta and the viewing angle
to the RS Oph system, we turn to the daily high-spectral-resolution
monitoring of the 2021 outburst of RS Oph presented by
\citet{2021arXiv210901101M, 2022arXiv220301378M} and use it to derive the
radial velocity of the leading edge of the ejecta.  At the very beginning of
the outburst, the glare of the RG wind ionised by the initial UV flash
overwhelmed the emission from the expanding ejecta.  Only by approximately
day +2 had recombination in the wind proceeded enough to allow a clean view
of the expanding ejecta.  Figure~\ref{fig:fig4} presents the H$\alpha$
profile of RS Oph for day +2.3.  Very sharp emission and absorption spikes
are located close to 0.0 radial velocity on the day +2.3 spectrum: the spike
in emission (FWHM$\sim$31~km\,s$^{-1}$,
RV$_\odot$=$-$6.3$\pm$0.4~km\,s$^{-1}$) originates from the recombining RG
wind, and that in absorption (FWHM$\sim$20~km\,s$^{-1}$,
RV$_\odot$=$-$64.5$\pm$0.8~km\,s$^{-1}$) comes from the outer and neutral
portion of the RG wind (its velocity matching the sum of the barycentric
velocity of RS~Oph, $-$40~km\,s$^{-1}$, and a 20-25~km\,s$^{-1}$ value for
the terminal velocity of the RG wind).  The fastest feature on the spectrum
for day +2.3 in Figure~\ref{fig:fig4} is the $-$4700~km\,s$^{-1}$ edge of
the P-Cyg absorption, tracing the fastest material external to the
pseudo-photosphere forming in the expanding, cooling, and recombining
ejecta.

At the time of our radio map, the H$\alpha$ profile of RS Oph has greatly
changed, as illustrated by the spectrum for day +34.2 in
Figure~\ref{fig:fig4}.  At that time, the bulk of the ejecta has been
significantly slowed down by the pre-existing RG wind, and the dominant
component of the H alpha emission has narrowed and reduced in both peak and
total intensity.  However, a broad pedestal persists ---highlighted by the
fitting in Figure~\ref{fig:fig4}--- that originates from the ejecta expelled
at the highest inclinations from the equatorial plane, and which suffered
from the smallest deceleration thanks to the reduced column density of the
RG wind along that direction.  On day +34.2, the broad pedestal was still
extending up to $-$4200~km\,s$^{-1}$.  Given the similarity of these two
values, we assume a linear deceleration with time, leading to $v_{\rm eje}
\cos i$=4450~km\,s$^{-1}$ as the mean radial velocity of the leading ejecta
during the interval from day 0 to +34.3 , where $i$ is the orbital
inclination (0$^\circ$ face-on, 90$^\circ$ edge-on).

The tangential velocity of the leading edge of the radio lobes in
Figure~\ref{fig:fig1}, averaged over the 34.3 days elapsed since the onset
of the outburst, is $v_{\rm eje}\sin i$$\sim$6100~km\,s$^{-1}$ at the {\it
Gaia} distance to RS Oph.  By comparing with the results from radial
velocities, we derive $v_{\rm eje}$=7550($\pm$150) km\,s$^{-1}$ for the
space velocity of the ejecta, and $i$=54$^\circ$($\pm$1) for the orbital
inclination.

It is interesting to compare the angular extent of the ejecta on our radio
map and on the [OIII]~5007 {\it HST} imaging of the 2006 ejecta presented by
\citet{2009ApJ...703.1955R}.  By day +155, the ejecta had travelled a total
east--west extension of 360$\pm$30 mas on HST images, which corresponds to
$v_{\rm eje}\sin i$$\sim$5400($\pm$500)~km\,s$^{-1}$ at the {\it Gaia}
distance.  This is very close to our $v_{\rm eje}\sin
i$$\sim$6100~km\,s$^{-1}$ estimated from the radio imaging in
Figure~\ref{fig:fig1}, especially considering that ($a$) some further
deceleration may have occurred between days +34.3 and +155, and ($b$)
[OIII]~5007 forms as a consequence of ion recombination, which is favoured
by a larger electron density, which is easier to encounter at inner radii of
the ejecta in homologous expansion compared to the radio synchrotron
emission originating at their outer edge.  It may then be concluded that the
ejection velocity in both 2006 and 2021 eruptions was very similar, and that
the orbital phase (the RG transiting at the ascending quadrature in 2006 and
at the descending one in 2021) made no difference.

At about day +22/+23  into the outburst, permitted lines began showing a
triple-peaked profile that progressively grew in contrast relative to the
much wider, underlying broad component.  HeI 6678 on the day +34.2 spectrum
of Figure~\ref{fig:fig4} shows an example of such a profile (for Balmer
lines, it only became visible at much later epochs when their strong broad
component weakened considerably).  The time of appearance of the
triple-peaked profile broadly coincides with the emergence of supersoft
X-ray emission \citep{2022MNRAS.514.1557P}, indicating that the
triple-peaked profile originates from the inner regions of RS~Oph, which
became visible only after the ejecta had become sufficiently diluted to
clear the view to the central binary.  The radial velocity of the outer
peaks declined from $\sim$$|$250$|$~km\,s$^{-1}$ at appearance to
$\sim$$|$150$|$~km\,s$^{-1}$ by day +102 when RS Oph entered the solar
conjunction \citep{2022arXiv220301378M}.

We believe that the outer peaks of the triple-peaked profiles originate from
the same location as the {\it CC} source in Figure~\ref{fig:fig1}; that is,
from the ring where the nova ejecta impact the DEOP and are more severely
decelerated by its greater density than by the diffuse RG wind at higher
latitudes.  Such a high-density location is also supported by the fact that
---contrary to permitted lines--- the forbidden nebular lines did not
develop a tripled peak profile (at least all those characterised by low
critical densities for collisional de-excitation, such as the usual [OIII]
or [NII]).  Assuming that the projected expansion velocity of the {\it CC}
ring declines exponentially from an initial 6100~km\,s$^{-1}$ to
250~km\,s$^{-1}$ on day +22 and 150~km\,s$^{-1}$ on day +102, by day +34.3
of our EVN image the ring has expanded to a diameter of 11~mas (14.5 AU),
which is compatible with the unresolved appearance of the {\it CC} in
Figure~\ref{fig:fig1}.

Finally, the central peak of the triple-peaked profiles remained relatively
sharp (FWHM $\sim$55 ~km\,s$^{-1}$) and its radial velocity stayed close to
the barycentric velocity of the central binary, suggesting that it may have
formed on the side of the RG irradiated by the still burning WD.

\section{Conclusions}

We present EVN radio imaging at 5 GHz of the expanding ejecta of the
recurrent nova RS Oph for day +34 of its 2021 outburst, extending for about
90~mas along the east--west direction.  Comparison with {\it Gaia} DR3
astrometry, radial velocity information from high-resolution optical
spectra, and similar radio observations of the 2006 eruption allows us to
combine all the observed features into a consistent 3D model.  The presence
of a strong density enhancement on the orbital plane ---seen at
$i$=54$^\circ$ inclination--- confined the nova ejecta primarily into two
lobes, expanding in opposite directions perpendicular to the plane at
$v_{\rm eje}$=7550 km\,s$^{-1}$ space velocity, averaged over the first 34
days.  The density enhancement partially obscures the receding lobe in the
background, and its inner radius impacted by the ejecta is the location of
the unresolved and bright central radio spike.

\begin{acknowledgements}

We acknowledge the rapid and constructive report by the referee (Stewart
Eyres).  Scientific results from data presented in this publication are
derived from the following EVN project code: RG012E.  The European VLBI
Network is a joint facility of independent European, African, Asian, and
North American radio astronomy institutes.  \textit{e}-MERLIN is a National
Facility operated by the University of Manchester at Jodrell Bank
Observatory on behalf of STFC, part of UK Research and Innovation.  BM
acknowledges financial support from the State Agency for Research of the
Spanish Ministry of Science and Innovation under grant PID2019-105510GB-C31
and through the Unit of Excellence Mar\'ia de Maeztu 2020--2023 award to the
Institute of Cosmos Sciences (CEX2019-000918-M).  This work has made use of
data from the European Space Agency (ESA) mission {\it Gaia}
(\url{https://www.cosmos.esa.int/gaia}), processed by the {\it Gaia} Data
Processing and Analysis Consortium (DPAC,
\url{https://www.cosmos.esa.int/web/gaia/dpac/consortium}).  Funding for the
DPAC has been provided by national institutions, in particular the
institutions participating in the {\it Gaia} Multilateral Agreement.

\end{acknowledgements}

\bibliographystyle{aa} 
\bibliography{paper.bib} 

\end{document}